\documentclass[11pt]{article}
\usepackage{amssymb}

\usepackage{amsmath}
\usepackage{theorem}
\usepackage{here}
\usepackage[dvipdfmx]{color}

\newtheorem{thm}{Theorem}[section]
\newtheorem{lem}{Lemma}[section]

\theorembodyfont{\rmfamily}
\newtheorem{algo}{Algorithm}[section]
\makeatletter
\renewcommand{\theequation}{%
   \thesection.\arabic{equation}}
\@addtoreset{equation}{section}
\makeatother

\usepackage{comment} %
\usepackage{bm}
\usepackage[pdftex]{graphicx}

\def\vc{{\check v}}

\def\Var{{\rm Var}}

\def\be{{\beta}}
\def\ga{{\gamma}}
\def\de{{\delta}}
\def\ep{{\varepsilon}}
\def\la{{\lambda}}

\def\ka{{\kappa}}
\def\ze{{\zeta}}

\def\bbe{{\text{\boldmath $\beta$}}}
\def\bga{{\text{\boldmath $\gamma$}}}

\def\bmu{{\text{\boldmath $\mu$}}}
\def\bnu{{\text{\boldmath $\nu$}}}

\def\bbet{{\widetilde \bbe}}

\def\De{{\Delta}}

\def\Ga{{\Gamma}}
\def\Om{{\Omega}}

\def\La{{\Lambda}}

\def\bOm{{\text{\boldmath $\Om$}}}

\def\bPsi{{\text{\boldmath $\Psi$}}}

\def\bPsih{{\widehat \bPsi}}

\def\b{{\text{\boldmath $b$}}}
\def\c{{\text{\boldmath $c$}}}

\def\n{{\text{\boldmath $n$}}}

\def\t{{\text{\boldmath $t$}}}
\def\u{{\text{\boldmath $u$}}}
\def\v{{\text{\boldmath $v$}}}

\def\x{{\text{\boldmath $x$}}}
\def\y{{\text{\boldmath $y$}}}

\def\I{{\text{\boldmath $I$}}}

\def\O{{\text{\boldmath $O$}}}

\def\W{{\text{\boldmath $W$}}}
\def\X{{\text{\boldmath $X$}}}

\def\ch{{\hat c}}

\def\uh{{\hat u}}
\def\vh{{\hat v}}

\def\ubh{{\hat \u}}
\def\vbh{{\hat \v}}

\def\vbt{{\tilde \v}}

\def\tr{{\rm tr\,}}

\def\[{{\text{\boldmath $[$}}}
\def\]{{\text{\boldmath $]$}}}

\def\/{{\Bigr/\!\!}}

\def\1r{{\rm (1)}}
\def\2r{{\rm (2)}}
\def\3r{{\rm (3)}}
\def\4r{{\rm (4)}}
\def\5r{{\rm (5)}}

\def\non{{\nonumber}}
\DeclareMathOperator*{\bdiag}{{\bf Diag\,}}

\begin{document}
\title{An Approximate Identity Link Function for Bayesian Generalized Linear Models}
\author{
Yasuyuki Hamura\footnote{Graduate School of Economics, Kyoto University, 
Yoshida-Honmachi, Sakyo-ku, Kyoto, 606-8501, JAPAN. 
\newline{
E-Mail: yasu.stat@gmail.com}} \
}
\maketitle
\begin{abstract}
In this note, we consider using a link function that has heavier tails than the usual exponential link function. 
We construct efficient Gibbs algorithms for Poisson and Multinomial models based on this link function by introducing gamma and inverse Gaussian latent variables and show that the algorithms generate geometrically ergodic Markov chains in simple settings. 
Our algorithms can be used for more complicated models with many parameters. 
We fit our simple Poisson model to a real dataset and confirm that the posterior distribution has similar implications to those under the usual Poisson regression model based on the exponential link function. 
Although less interpretable, our models are potentially more tractable or flexible from a computational point of view in some cases. 

\par\vspace{4mm}
\end{abstract}

\section{Introduction}
\label{sec:introduction}
In constructing generalized linear models, the exponential link function is widely used to relate positive-valued parameters to underlying real-valued variables as the simplest and interpretable default choice leading to reasonable results in most applications; see, for example, McCullagh and Nelder (1989). 
Although the exponential link function is good in practice, it is not necessarily desirable from a Bayesian computational point of view. 
For instance, in the Poisson and generalized gamma regression cases, posterior densities of regression coefficients are expressed using the double exponential function, which makes most priors nonconjugate, and we will have to resort to some specialized algorithms such as the Metropolis--Hastings algorithms of D'Angelo and Canale (2023) and Shukla, Ranjan and Upadhyay (2023) or the rejection sampler (e.g., Doss and Narasimhan (1994)). 
Also, the data augmentation approach has been proved generally useful in generating posterior samples but does not seem so in the present case, where the target density already has light tails and is difficult to express as a marginal density or a sum of more light-tailed densities. 

We note that the exponential link function is not the only link function to be used. 
For example, in the binomial case, the probit function can be used instead of the logit function, which is based on the exponential function. 
Wang and Dey (2010) generalized the complementary log-log function for modeling binary data. 
Komaki (2024) introduced a Poisson regression model based on a quadratic link function with a theoretically natural structure. 

In this paper, we consider using a link function, in place of the exponential function, to contruct models that are potentially more tractable or flexible from a computational point of view in some cases, at the cost of lower interpretability of regression coefficients. 
One feature of our link function is that it is monotone in its argument. 
Because of this feature, our models are supposed to retain some interpretability. 
Another feature of our link function is that it is asymptotically equivalent to the identity link function. 
Because of this, posterior distributions of regression coefficients have heavier tails than those under usual models based on the exponential link function. 
We can construct Gibbs samplers by introducing gamma and inverse Gaussian latent variables, both of which are easily sampled. 

Some preliminary results are given in Section \ref{subsec:results}. 
In Sections \ref{sec:Po} and \ref{sec:Multin}, respectively, we consider Poisson and Multinomial regression models based on our link function. 
In particular, we prove geometric ergodicity of the Markov chain based on each algorithm to theoretically show that it is efficient. 
Although Choi and Hobert (2013) established uniform ergodicity for the famous Polya-Gamma algorithm of Polson, Scott and Windle (2013) for logistic models, efficiency of algorithms for Poisson regression models have not been fully studied as far as the author knows.

\subsection{Preliminaries}
\label{subsec:results} 
\begin{figure}[!htb]
\centering
\includegraphics[width = \linewidth]{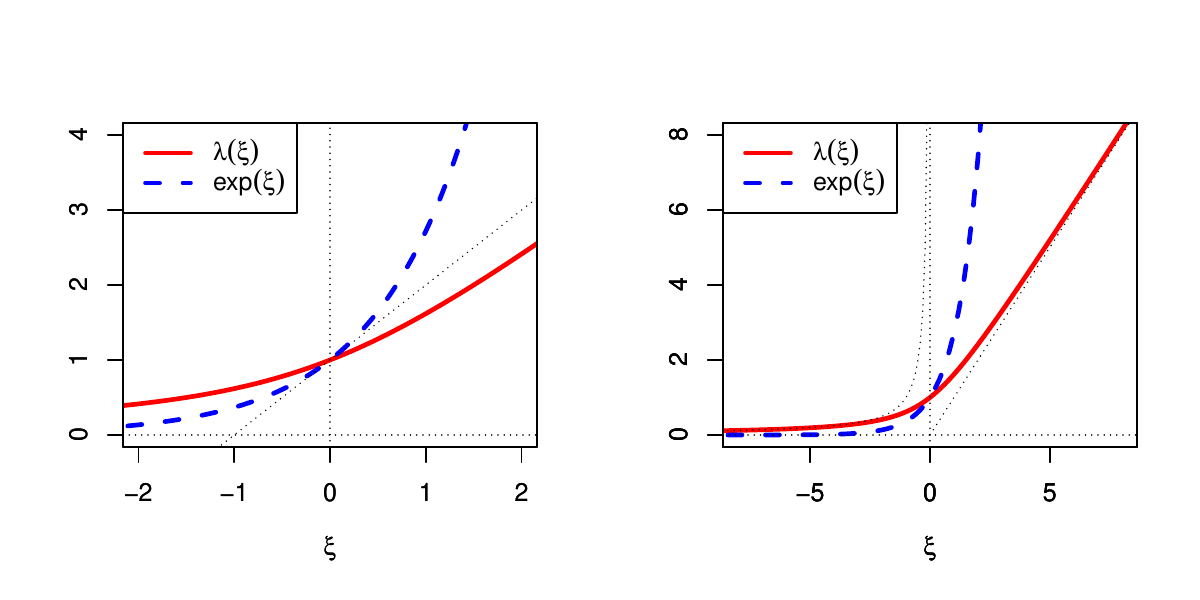}
\caption{The link function $\la $ given by (\ref{eq:link}) and the exponential link function. }
\label{fig:curve} 
\end{figure}

The inverse of the diffeomorphism $(0, \infty ) \ni u \mapsto u - 1 / u \in \mathbb{R}$ is 
\begin{align}
&\la \colon \mathbb{R} \ni \xi \mapsto \la ( \xi ) = {\xi + \sqrt{\xi ^2 + 4} \over 2} = {2 \over - \xi + \sqrt{\xi ^2 + 4}} \in (0, \infty ) \text{,} \label{eq:link} 
\end{align}
which is monotonically increasing. 
It can be seen that $\la ( \xi ) \sim | \xi |^{\pm 1}$ as $\xi \to \pm \infty $. 
We consider using this function in place of the famous exponential link function. 
These two functions are shown in Figure \ref{fig:curve}.

In order to construct a Gibbs sampler, we use the fact that $e^{- \ka }$ is the normalizing constant of an  inverse Gaussian distribution, 
\begin{align}
e^{- \ka } %
&= \int_{0}^{\infty } {1 \over (2 \pi )^{1 / 2}} {1 \over \ze ^{3 / 2}} e^{- 1 / (2 \ze )} e^{- ( \ze/ 2) \ka ^2} d\ze \text{,} \label{eq:DE} 
\end{align}
for any $\ka > 0$, as in Park and Casella (2008). 
For $\mu , \la > 0$, the inverse Gaussian distribution with parameters $( \mu , \la )$ has density given by 
\begin{align}
{\rm{IGauss}} (v | \mu , \la ) &= \sqrt{{\la \over 2 \pi }} \exp \Big( {\la \over \mu } \Big) {1 \over v^{3 / 2}} \exp \Big\{ - {1 \over 2} \Big( {\la \over \mu ^2} v + {\la \over v} \Big) \Big\} \text{,} \quad v \in (0, \infty ) \text{.} \non 
\end{align}

\section{The Poisson Case}
\label{sec:Po} 
\subsection{The model and posterior sampling}
\label{subsec:model} 
We illustrate the use of the link function (\ref{eq:link}) in a simple Poisson case. 
Let $n, p \in \mathbb{N}$ and suppose that 
\begin{align}
\bbe &\sim {\rm{N}}_p ( \bbe | \bmu , \bPsi ^{- 1} ) \propto \exp \Big\{ - {1 \over 2} ( \bbe - \bmu )^{\top } \bPsi ( \bbe - \bmu ) \Big\} \text{.} \non 
\end{align}
for $\bmu \in \mathbb{R} ^p$ and $\bPsi > \O ^{(p)}$ and that 
\begin{align}
y_i &\sim {\rm{Po}} \Big( y_i \Big| n_i {\be _i + \sqrt{{\be _i}^2 + 4} \over 2} \Big) = {{n_i}^{y_i} \over y_i !} \Big( {\be _i + \sqrt{{\be _i}^2 + 4} \over 2} \Big) ^{y_i} \exp \Big( - n_i {\be _i + \sqrt{{\be _i}^2 + 4} \over 2} \Big) \text{,} \non 
\end{align}
where $\be _i = {\x _i}^{\top } \bbe $, for known $\x _i \in \mathbb{R} ^p$ and $n_i > 0$ for $i = 1, \dots , n$. 
The observable outcome and explanatory variables are $\y = ( y_i )_{i = 1}^{n}$ and $\X = ( \x _1 , \dots , \x _n )^{\top }$, whereas the unobservable parameter vector of regression coefficients are $\bbe $. 
Thus, the mean function of $y_i$ is given by $n_i \la ( {\x _i}^{\top } \bbe )$ for $i = 1, \dots , n$. 

It follows from (\ref{eq:DE}) that for all $i = 1, \dots , n$, 
\begin{align}
p( y_i | \bbe ) &= {(2 n_i )^{y_i} \over y_i !} {1 \over (- \be _i + \sqrt{{\be _i}^2 + 4} )^{y_i}} \exp \{ - ( n_i / 2) ( \be _i + \sqrt{{\be _i}^2 + 4} ) \} \non \\
&= {(2 n_i )^{y_i} \over y_i !} \int_{0}^{\infty } {{u_i}^{y_i - 1} \over \Ga ( y_i )} \exp \Big\{ - \Big( {n_i \over 2} - u_i \Big) \be _i \Big\} \exp \Big\{ - \Big( {n_i \over 2} + u_i \Big) \sqrt{{\be _i}^2 + 4} \Big\} d{u_i} \non \\
&= {(2 n_i )^{y_i} \over y_i !} \int_{(0, \infty )^2} {{u_i}^{y_i - 1} \over \Ga ( y_i )} \exp \Big\{ - \Big( {n_i \over 2} - u_i \Big) \be _i \Big\} \non \\
&\quad \times {1 \over (2 \pi )^{1 / 2}} {1 \over {v_i}^{3 / 2}} e^{- 1 / (2 v_i )} \exp \Big\{ - {v_i \over 2} \Big( {n_i \over 2} + u_i \Big) ^2 ( {\be _i}^2 + 4) \Big\} d( u_i , v_i ) \non 
\end{align}
if $y_i \ge 1$ and 
\begin{align}
p( y_i | \bbe ) &= \exp \{ - ( n_i / 2) ( \be _i + \sqrt{{\be _i}^2 + 4}) \} \non \\
&= \int_{0}^{\infty } \exp \Big( - {n_i \over 2} \be _i \Big) {1 \over (2 \pi )^{1 / 2}} {1 \over {v_i}^{3 / 2}} e^{- 1 / (2 v_i )} \exp \Big\{ - {v_i \over 2} \Big( {n_i \over 2} \Big) ^2 ( {\be _i}^2 + 4) \Big\} d{v_i} \non 
\end{align}
and $u_i = 0$ if $y_i = 0$. 
Thus, we can use the following Gibbs sampler. 
\begin{algo}
\label{algo:Po} 
The regression coefficients $\bbe $ are updated in the following way. 
\begin{itemize}
\item
For each $i = 1, \dots , n$, 
\begin{itemize}
\item
Sample $u_i \sim {\rm{Ga}} ( u_i | y_i , - \be _i + \sqrt{{\be _i}^2 + 4} )$, 
\item
sample $v_i \sim {\rm{IGauss}} ( v_i | 1 / \{ ( n_i / 2 + u_i ) \sqrt{{\be _i}^2 + 4} \} , 1)$, and 
\item
let $w_i = v_i ( n_i / 2 + u_i )^2$. 
\end{itemize}
\item
Let $\W = \bdiag ( w_i )_{i = 1}^{n}$ and sample 
\begin{align}
\bbe &\sim {\rm{N}}_p ( \bbe | ( \bPsi + \X ^{\top } \W \X )^{- 1} \{ \bPsi \bmu + \X ^{\top } ( \u - \n / 2) \} , ( \bPsi + \X ^{\top } \W \X )^{- 1} ) \text{,} \non 
\end{align}
where $\u = ( u_i )_{i = 1}^{n}$ and $\n = ( n_i )_{i = 1}^{n}$. 
\end{itemize}
\end{algo}

\subsection{Geometric ergodicity}
\label{subsec:ergodicity} 
Algorithm \ref{algo:Po} can be theoretically shown to be efficient. 
For technical terms such as geometric ergodicity, see Jones and Hobert (2001). 
In this section, we fix $\y \in {\mathbb{N} _0}^n$. 
For simplicity, we assume that all observations are strictly positive. 
\begin{thm}
\label{thm:Po} 
Assume that $y_i \ge 1$ for all $i = 1, \dots , n$. 
Suppose that $\X ^{\top } \X $ is of full rank. 
Then the Markov chain based on Algorithm \ref{algo:Po} is geometrically ergodic. 
\end{thm}

In order to prove the above result, we use the drift and minorization technique and in particular the energy function 
\begin{align}
&V \colon \mathbb{R} ^p \ni \bbe \mapsto V( \bbe ) = \bbe ^{\top } \bPsi \bbe \in [0, \infty ) \text{.} \label{eq:drift} 
\end{align}
This is similar to the energy function of Roy and Hobert (2010), for example, and not to that of Bhattacharya, Khare and Pal (2022), who considered a data augmentation scheme for the prior rather than the likelihood. 
Let 
\begin{align}
k( \bbe | \bbe ^{\rm{o}} ) &= \int_{(0, \infty )^n \times (0, \infty )^n} p^{\bbe | ( \u , \v , \y )} ( \bbe | \u , \v , \y ) p^{( \u , \v ) | ( \bbe , \y )} ( \u , \v | \bbe ^{\rm{o}} , \y ) d( \u , \v ) \non 
\end{align}
for $\bbe , \bbe ^{\rm{o}} \in \mathbb{R} ^p$ and let 
\begin{align}
(P V) ( \bbe ^{\rm{o}} ) &= E^{( \u , \v ) | ( \bbe , \y )} [ E^{\bbe | ( \u , \v , \y )} [ V( \bbe ) | \u , \v , \y ] | \bbe ^{\rm{o}} , \y ] \non 
\end{align}
for $\bbe ^{\rm{o}} \in \mathbb{R} ^p$. 

As in Roy and Hobert (2010), it is straightforward to check the minorization condition associated with (\ref{eq:drift}) for $k( \bbe | \bbe ^{\rm{o}} )$; see Lemma \ref{lem:minorization}. 
In verifying the drift condition that there exist $0 < \de < 1$ and $\De > 0$ such that $(P V) ( \bbe ) \le \De + \de V( \bbe )$ for all $\bbe \in \mathbb{R} ^p$, %
we basically estimate $(P V) ( \bbe )$ considering the limit as $\| \bbe \| \to \infty $, since near the origin, it is bounded by some constant (Lemma \ref{lem:bound}). 

Because the hyperparameters $( \bmu , \bPsi )$ and $\n $ are fixed, our analysis is not practical. 
Theorem \ref{thm:Po} reassures us that the Markov chain based on the data augmentation scheme is at least geometrically ergodic in the simplest setting. 
Additionally, although the proof depends heavily on the condition that $\bPsi > \O ^{(p)}$ is fixed, some ideas may be useful in other cases.

\subsection{A numerical example}
Algorithm \ref{algo:Po} was applied to data considered in Chapter 10 of Hoff (2009). 
Here, $(n, p) = (52, 3)$ and, for each sparrow $i = 1, \dots , n$, the outcome $y_i$ is the number of offspring and $x_{i, 2}$ is the age, with $x_{i, 1} = 1$, $x_{i, 3} = {x_{i, 2}}^2$, and $n_i = 1$. 
For comparison, we also fit the usual Poisson regression model where 
\begin{align}
&y_i \sim {\rm{Po}} ( y_i | \exp ( {\x _i}^{\top } \bga )) \text{,} \quad i = 1, \dots , n \text{,} \non \\
&\bga = ( \ga _k )_{k = 1}^{3} \sim {\rm{N}}_p ( \bga | \bmu _{\bga } , {\bPsi _{\bga }}^{- 1} ) \non 
\end{align}
using the R package \verb|bpr| of D'Angelo and Canale (2023). 
We set $\bmu = \bmu _{\bga } = (0, 0, 0)^{\top }$ and ${\bPsi }^{- 1} = {\bPsi _{\bga }}^{- 1} = \bdiag (100, 100, 100)$. 
We generated $5,000$ samples after discarding $5,000$ samples as burn-in. 

\begin{figure}[H]%
\centering
\includegraphics[width = \linewidth]{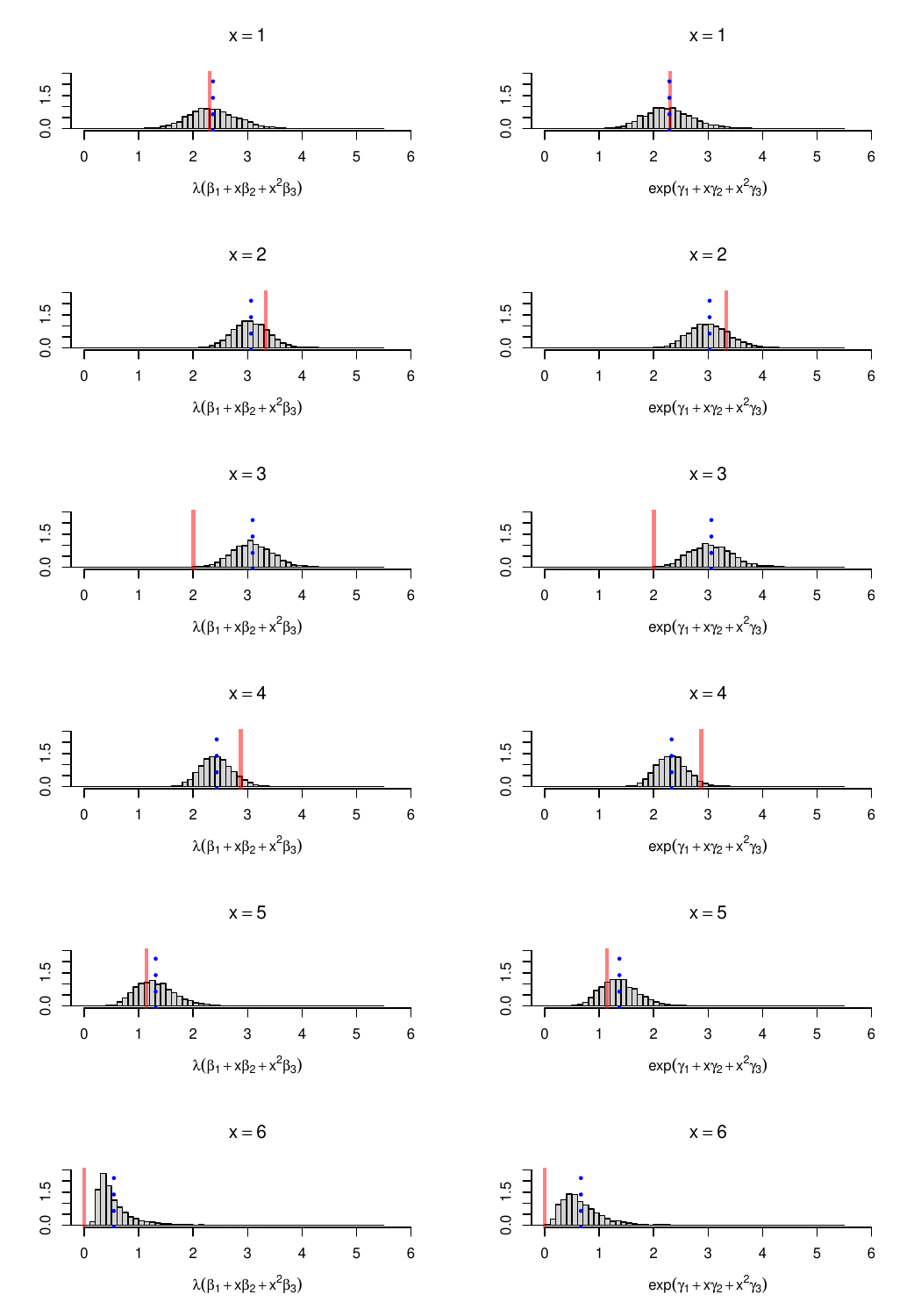}
\caption{Posterior predictive distributions of the conditional means %
for the sparrow data. 
The left and right panels correspond to our model and the usual models, respectively. 
Each row corresponds to each age ranging from $1$ to $6$. 
The dotted blue lines show the expectations of the predictive distributions. 
The solid red lines show the mean of $\{ y_i | x_{i, 2} = x \} $ for $x = 1, \dots , 6$. }
\label{fig:sparrow} 
\end{figure}

Posterior distributions of the conditional means $\la ((1, x, x^2 ) \bbe )$ and $\exp ((1, x, x^2 ) \bga )$ under the two models are shown in Figure \ref{fig:sparrow}, where $x \in \{ 1, \dots , 6 \} $. 
It is clear that our model generated simliar samples to those under the usual model. 
Our model lacks the interpretability of the parameters on the logarithmic scale but seems practically reasonable in predictive terms. 
The only discernible difference would be in the sixth row, where the left density is more peaked than the right. 
This case almost corresponds to out-of-sample prediction since the numbers of sparrows in the sample corresponding to $x = 1, \dots , 6$ are $10, 9, 9, 16, 7, 1$, respectively. 

We remark that the algorithm of D'Angelo and Canale (2023) genereted more efficient Markov chains, possibly reflecting some structural difference between the models or algorithms. 
However, Markov chains generated by Algorithm \ref{algo:Po} were not very inefficient, as Theorem \ref{thm:Po} suggests, at least compared with Figure 10.5 of Hoff (2009). 
Our simple algorithm without any tuning parameters could potentially be useful in more complicated settings where computation time would also be important.

\subsection{Reproductive properties}
Here, we briefly note that we can implicitly use the reproductive properties of the latent variables introduced in Section \ref{subsec:model}. 
Suppose that $\x _1 = \x _2$ with $n \ge 2$. 
Then 
\begin{align}
p( y_1 , y_2 | \bbe ) &\stackrel{\rm{\bbe }}{\propto } \{ \la ( {\x _1}^{\top } \bbe ) \} ^{y_1 + y_2} \exp \{ - ( n_1 + n_2 ) \la ( {\x _1}^{\top } \bbe ) \} \text{.} \non 
\end{align}
Therefore, we could sample $\tilde{u} \sim {\rm{Ga}} ( y_1 + y_2 , - \be _1 + \sqrt{{\be _1}^2 + 4} )$ and $\tilde{v} \sim {\rm{IGauss}} (1 / \{ ( n_1 / 2 + n_2 / 2 + \tilde{u} ) \sqrt{{\be _1}^2 + 4} \} , 1)$ instead of $( u_1 , v_1 )$ and $( u_2 , v_2 )$. 
However, the two approaches would be equaivalent since, by the reproductive properties of the gamma and inverse Gaussian distributions, 
\begin{align}
&( \x _1 u_1 + \x _2 u_2 ) | \bbe \stackrel{\rm{d}}{=} ( \x _1 \tilde{u} ) | \bbe \quad \text{and} \non \\
&( \x _1 w_1 {\x _1}^{\top } + \x _2 w_2 {\x _2}^{\top } ) | ( u_1 , u_2 , \bbe ) \stackrel{\rm{d}}{=} ( \x _1 \tilde{w} {\x _1}^{\top } ) | (( \tilde{u} = u_1 + u_2 ), \bbe ) \text{,} \non 
\end{align}
where $\tilde{w} = \tilde{v} ( n_1 / 2 + n_2 / 2 + \tilde{u} )^2$, and since the conditional distribution of $\bbe $ depends on $( u_1 , u_2 )$ only through the left-hand sides of the above equations.

\section{The Multinomial Case}
\label{sec:Multin} 
In this section, we consider a multinomial regression model. 
Let $n \in \mathbb{N}$ and let $m_i , p_i  \in \mathbb{N}$ for $i = 1, \dots , n$. 
Suppose that $\bbe \sim {\rm{N}}_p ( \bbe | \bmu , \bPsi ^{- 1} )$ and that 
\begin{align}
\y _i = ( y_{i, k} )_{k = 0}^{p_i} &\sim {\rm{Multin}}_{p_i} \Big( \y _i \Big| m_i , \Big( {\la _{i, k} \over 1 + \La _i} \Big) _{k = 1}^{p_i} \Big) \non \\
&= {m_i ! \over \prod_{k = 0}^{p_i} y_{i, k} !} {1 \over (1 + \La _i )^{y_{i, 0}}} \prod_{k = 1}^{p_i} \Big( {\la _{i, k} \over 1 + \La _i} \Big) ^{y_{i, k}} \text{,} \non 
\end{align}
where $\x _{i, k} \in \mathbb{R} ^p$, $\be _{i, k} = {\x _{i, k}}^{\top } \bbe $, and $\la _{i, k} = \la ( \be _{i, k} )$ for $k = 1, \dots , p_i$ and $\La _i = \sum_{k = 1}^{p_i} \la _{i, k}$, for $i = 1, \dots , n$. 
This is a multinomial reponse model of section 6.4.2 of McCullagh and Nelder (1989) with the exponential link function replaced by $\la $. 
The polychotomous regression of Holmes and Held (2006) is also covered. 

For all $i = 1, \dots , n$, we have 
\begin{align}
p( \y _i | \bbe ) &\stackrel{\bbe }{\propto } {1 \over (2 + 2 \La _i )^{m_i}} \prod_{k = 1}^{p_i} {1 \over (- \be _{i, k} + \sqrt{{\be _{i, k}}^2 + 4} )^{y_{i, k}}} \non \\
&= \int_{(0, \infty )^{1 + p_i}} \Big( {{u_{i, 0}}^{m_i - 1} e^{- 2 u_{i, 0}} \over \Ga ( m_i )} \prod_{k = 1}^{p_i} \Big[ \exp \{ - ( \be _{i, k} + \sqrt{{\be _{i, k}}^2 + 4} ) u_{i, 0} \} \non \\
&\quad \times {{u_{i, k}}^{y_{i, k} - 1} \over \Ga ( y_{i, k} )} \exp \{ - (- \be _{i, k} + \sqrt{{\be _{i, k}}^2 + 4} ) u_{i, k} \} \Big] \Big) d{\u _i} \non \\
&= \int_{(0, \infty )^{1 + p_i}} \Big( \Big\{ {{u_{i, 0}}^{m_i - 1} e^{- 2 u_{i, 0}} \over \Ga ( m_i )} \prod_{k = 1}^{p_i} {{u_{i, k}}^{y_{i, k} - 1} \over \Ga ( y_{i, k} )} \Big\} \non \\
&\quad \times \prod_{k = 1}^{p_i} [ \exp \{ \be _{i, k} ( u_{i, k} - u_{i, 0} ) \} \exp \{ - ( u_{i, k} + u_{i, 0} ) \sqrt{{\be _{i, k}}^2 + 4} \} ] \Big) d{\u _i} \non \\
&= \int_{(0, \infty )^{1 + p_i}} \Big( \Big\{ {{u_{i, 0}}^{m_i - 1} e^{- 2 u_{i, 0}} \over \Ga ( m_i )} \prod_{k = 1}^{p_i} {{u_{i, k}}^{y_{i, k} - 1} \over \Ga ( y_{i, k} )} \Big\} \prod_{k = 1}^{p_i} \Big[ \exp \{ \be _{i, k} ( u_{i, k} - u_{i, 0} ) \} \non \\
&\quad \times \int_{0}^{\infty } {1 \over (2 \pi )^{1 / 2}} {1 \over {v_{i, k}}^{3 / 2}} e^{- 1 / (2 v_{i, k} )} \exp \Big\{ - {v_{i, k} \over 2} ( u_{i, k} + u_{i, 0} )^2 ( {\be _{i, k}}^2 + 4) \Big\} \Big] d{v_{i, k}} \Big) d{\u _i} \text{.} \non 
\end{align}
Therefore, 
\begin{align}
&p( \v | \u , \bbe , \y ) = \prod_{i = 1}^{n} \prod_{k = 1}^{p_i} {\rm{IGauss}} ( v_{i, k} | 1 / \{ ( u_{i, k} + u_{i, 0} ) \sqrt{{\be _{i, k}}^2 + 4} \} , 1) \text{,} \non \\
&p( \u | \bbe , \y ) = \prod_{i = 1}^{n} \Big\{ {\rm{Ga}} ( u_{i, 0} | m_i , 2 + 2 \La _i ) \prod_{k = 1}^{p_i} {\rm{Ga}} ( u_{i, k} | y_{i, k} , - \be _{i, k} + \sqrt{{\be _{i, k}}^2 + 4} ) \Big\} \text{,} \non \\
&p( \bbe | \v , \u , \y ) \propto {\rm{N}}_p \Big( \bbe \Big| \bPsih ^{- 1} \Big\{ \bPsi \bmu + \sum_{i = 1}^{n} \sum_{k = 1}^{p_i} \x _{i, k} ( u_{i, k} - u_{i, 0} ) \Big\} , \bPsih ^{- 1} \Big) \text{,} \non 
\end{align}
where 
\begin{align}
\bPsih &= \bPsi + \sum_{i = 1}^{n} \sum_{k = 1}^{p_i} \x _{i, k} v_{i, k} ( u_{i, k} + u_{i, 0} )^2 {\x _{i, k}}^{\top } \text{,} \non 
\end{align}
and we can use the following Gibbs sampler to sample $\bbe $. 
\begin{algo}
\label{algo:Multin} 
The regression coefficients $\bbe $ are updated in the following way. 
\begin{itemize}
\item
Sample $\u \sim p( \u | \bbe , \y )$ and then $\v \sim p( \v | \u , \bbe , \y )$. 
\item
Sample $\bbe \sim p( \bbe | \v , \u , \y )$. 
\end{itemize}
\end{algo}

As in Section \ref{subsec:ergodicity}, we consider geometric ergodicity. 
\begin{thm}
\label{thm:multin} 
Suppose that 
\begin{align}
&\sum_{i = 1}^{n} \sum_{k = 1}^{p_i} \x _{i, k} {\x _{i, k}}^{\top } > \O ^{(p)} \text{.} \non 
\end{align}
Suppose that $y_{i, k} \ge 1$ for all $k = 1, \dots , p_i$ for all $i = 1, \dots , n$. 
Then the Markov chain based on the Algorithm \ref{algo:Multin} is geometrically ergodic. 
\end{thm}

In the polychotomous regression case where $p_1 = \dots = p_n$ and $\be _{i, k} = {\x _i}^{\top } \bbe _k$ for some $\x _i , \bbe _k \in \mathbb{R} ^{p / p_i}$, $k = 1, \dots , p_i$, $i = 1, \dots , n$, one computational advantage of our approach is that we can simultaneously update all the elements of $\bbe $. 
In the usual setting based on the exponential link function, we would sequentially update each block of $\bbe $, which could have some undesirable effect on efficiency as well as computation time.

Finally, we remark that we can construct a more efficient algorithm in the Bernoulli case with $p_1 = \dots = p_n = 1$ and $m_1 = \dots = m_n = 1$. 
Fix $i = 1, \dots , n$ and suppose that $y_{i, 1} = 1$. 
Then 
\begin{align}
p( \y _i | \bbe ) = {m_i ! \over \prod_{k = 0}^{p_i} y_{i, k} !} \Big( {\la _{i, 1} \over 1 + \la _{i, 1}} \Big) ^{y_{i, 1}} \text{.} \non 
\end{align}
Therefore, only one set of latent variables will do since $\la _{i, 1} / (1 + \la _{i, 1} ) = 2 / \{ 2 + (- \be _{i, 1} + \sqrt{{\be _{i, 1}}^2 + 4} ) \} $, whereas two sets of latent variables are introduced for Algorithm \ref{algo:Multin}. 
However, this will be practically unimportant; the logistic regression model and the efficient P\'{o}lya-Gamma sampler of Polson, Scott and Windle (2013) will be used in this case.

\section*{Funding}
Research of the author was supported in part by JSPS KAKENHI Grant Number JP22K20132 from Japan Society for the Promotion of Science.

\newpage
\setcounter{page}{1}
\setcounter{equation}{0}
\renewcommand{\theequation}{S\arabic{equation}}
\setcounter{section}{0}
\renewcommand{\thesection}{S\arabic{section}}
\setcounter{table}{0}
\renewcommand{\thetable}{S\arabic{table}}
\setcounter{figure}{0}
\renewcommand{\thefigure}{S\arabic{figure}}

\begin{center}
{\LARGE\bf Supplementary Materials}
\end{center}

\bigskip

\section{Lemmas}
\label{sec:lemmas} 
Four lemmas used to prove Theorem \ref{thm:Po} are given in this section. 

\begin{lem}
\label{lem:minorization} 
Assume that $y_i \ge 1$ for all $i = 1, \dots , n$. 
Then for any $L > 0$, there exist a normalized density function $f$ on $\mathbb{R} ^p$ and $\ep > 0$ such that we have $k( \bbe | \bbe ^{\rm{o}} ) \ge \ep f( \bbe )$ for all $\bbe ^{\rm{o}} \in \{ \bbet \in \mathbb{R} ^p | V( \bbet ) \le L \} $ and all $\bbe \in \mathbb{R} ^p$. 
\end{lem}

\noindent
{\bf Proof%
.} \ \ Fix $L > 0$ and let $C = \{ \bbet \in \mathbb{R} ^p | \bbet ^{\top } \bPsi \bbet \le L \} $. 
Fix $\bbe ^{\rm{o}} \in C$. 

Fix $i = 1, \dots , n$ and let $\be _{i}^{\rm{o}} = {\x _i}^{\top } \bbe ^{\rm{o}}$. 
Then 
\begin{align}
p( u_i , v_i | \bbe ^{\rm{o}} , \y ) &= {( b_{i}^{\rm{o}} )^{y_i} \over \Ga ( y_i )} {u_i}^{y_i - 1} \exp (- b_{i}^{\rm{o}} u_i ) {1 \over \sqrt{2 \pi }} \exp \Big( {1 \over \mu _{i}^{\rm{o}}} \Big) {1 \over {v_i}^{3 / 2}} \exp \Big[ - {1 \over 2} \Big\{ {1 \over ( \mu _{i}^{\rm{o}} )^2} v_i + {1 \over v_i} \Big\} \Big] \text{,} \non 
\end{align}
where $b_{i}^{\rm{o}} = - \be _{i}^{\rm{o}} + \sqrt{( \be _{i}^{\rm{o}} )^2 + 4}$ and $\mu _{i}^{\rm{o}} = 1 / \{ ( n_i / 2 + u_i ) \sqrt{( \be _{i}^{\rm{o}} )^2 + 4} \} $. 
Since 
\begin{align}
&\mu _{i}^{\rm{o}} \le {1 \over n_i} \text{,} \quad {2 \over \sqrt{( \be _{i}^{\rm{o}} )^2 + 4}} \le b_{i}^{\rm{o}} \le 2 \sqrt{( \be _{i}^{\rm{o}} )^2 + 4} \text{,} \quad \text{and} \quad | \be _{i}^{\rm{o}} | \le \| \x _i \| \| \bbe ^{\rm{o}} \| \non 
\end{align}
and since $\| \bbe ^{\rm{o}} \| ^2 \le M_0 ( \bbe ^{\rm{o}} )^{\top } \bPsi \bbe ^{\rm{o}} \le M_0 L$, where $M_0$ is the reciprocal of the smallest eigenvalue of $\bPsi $, it follows that 
\begin{align}
&{1 \over n_i / 2 + u_i} {1 \over \sqrt{M_0 L \| \x _i \| ^2 + 4}} \le \mu _{i}^{\rm{o}} \le {1 \over n_i} \quad \text{and} \non \\
&{2 \over \sqrt{M_0 L \| \x _i \| ^2 + 4}} \le b_{i}^{\rm{o}} \le 2 \sqrt{M_0 L \| \x _i \| ^2 + 4} \text{.} \non 
\end{align}
Therefore, $p( u_i , v_i | \bbe ^{\rm{o}} , \y ) \ge q_i ( u_i, v_i )$, where 
\begin{align}
q_i ( u_i, v_i ) &= {1 \over \Ga ( y_i )} \Big( {2 \over \sqrt{M_0 L \| \x _i \| ^2 + 4}} \Big) ^{y_i} {u_i}^{y_i - 1} \exp (- 2 \sqrt{M_0 L \| \x _i \| ^2 + 4} u_i ) \non \\
&\quad \times {1 \over \sqrt{2 \pi }} \exp ( n_i ) {1 \over {v_i}^{3 / 2}} \exp \Big[ - {1 \over 2} \Big\{ ( n_i / 2 + u_i )^2 (M_0 L \| \x _i \| ^2 + 4) v_i + {1 \over v_i} \Big\} \Big] \non 
\end{align}
is a integrable function of $( u_i , v_i ) \in (0, \infty )^2$. 

We obtain 
\begin{align}
k( \bbe | \bbe ^{\rm{o}} ) &\ge \int_{(0, \infty )^n \times (0, \infty )^n} p( \bbe | \u , \v , \y ) \Big\{ \prod_{i = 1}^{n} q_i ( u_i, v_i ) \Big\} d( \u , \vbt ) \non 
\end{align}
for all $\bbe \in \mathbb{R} ^p$. 
The result follows. 
\hfill$\Box$

\begin{lem}
\label{lem:bound} 
For any $R > 0$, we have 
\begin{align}
&\sup_{\bbe ^{\rm{o}} \in \{ \bbet \in \mathbb{R} ^p | \| \bbet \| \le R \} } (P V)( \bbe ^{\rm{o}} ) < \infty \text{.} \non 
\end{align}
\end{lem}

\noindent
{\bf Proof%
.} \ \ Fix $R > 0$ and $\bbe ^{\rm{o}} \in \mathbb{R} ^p$. 
We have 
\begin{align}
(P V) ( \bbe ^{\rm{o}} ) &\le ( \tr \bPsi ) E^{( \u , \v ) | ( \bbe , \y )} [ E^{\bbe | ( \u , \v , \y )} [ \| \bbe \| ^2 | \u , \v , \y ] | \bbe ^{\rm{o}} , \y ] \non \\
&\le ( \tr \bPsi ) \sum_{k = 1}^{p} E^{( \u , \v ) | ( \bbe , \y )} [ ( \bPsi )^{k, k} \non \\
&\quad + ( E^{\bbe | ( \u , \v , \y )} [ \bbe | \u , \v , \y ])^{\top } [ \{ \tr ( \bPsi ^{- 1} ) \} \bPsi ] E^{\bbe | ( \u , \v , \y )} [ \bbe | \u , \v , \y ] | \bbe ^{\rm{o}} , \y ] \non \\
&\le ( \tr \bPsi ) \sum_{k = 1}^{p} E^{( \u , \v ) | ( \bbe , \y )} [ ( \bPsi )^{k, k} \non \\
&\quad + \{ \tr ( \bPsi ^{- 1} ) \} \{ \bPsi \bmu + \X ^{\top } ( \u - \n / 2) \} ^{\top } ( \bPsi + \X ^{\top } \W \X )^{- 1} \{ \bPsi \bmu + \X ^{\top } ( \u - \n / 2) \} | \bbe ^{\rm{o}} , \y ] \non \\
&\le 2 ( \tr \bPsi ) \sum_{k = 1}^{p} E^{( \u , \v ) | ( \bbe , \y )} [ ( \bPsi )^{k, k} + \{ \tr ( \bPsi ^{- 1} ) \} ( \bPsi \bmu )^{\top } \bPsi ^{- 1} ( \bPsi \bmu ) \non \\
&\quad + \{ \tr ( \bPsi ^{- 1} ) \} \{ \X ^{\top } ( \u - \n / 2) \} ^{\top } \bPsi ^{- 1} \{ \X ^{\top } ( \u - \n / 2) \} | \bbe ^{\rm{o}} , \y ] \text{,} \non 
\end{align}
where the third and fourth inequalities follow since $\bPsi \le \bPsi + \X ^{\top } \W \X $. 
Since $\| \bbe ^{\rm{o}} \| \le R$ implies that 
\begin{align}
&E^{( \u , \v ) | ( \bbe , \y )} [ {u_i}^2 | \bbe ^{\rm{o}} , \y ] = {y_i \over ( b_{i}^{\rm{o}} )^2} + {{y_i}^2 \over ( b_{i}^{\rm{o}} )^2} \le ( y_i + {y_i}^2 ) \Big\{ {\sqrt{( {\x _i}^{\top } \bbe ^{\rm{o}} )^2 + 4} \over 2} \Big\} ^2 \le {y_i + {y_i}^2 \over 4} ( R^2 \| \x _i \| ^2 + 4) \text{,} \non 
\end{align}
where $b_{i}^{\rm{o}} = - {\x _i}^{\top } \bbe ^{\rm{o}} + \sqrt{( {\x _i}^{\top } \bbe ^{\rm{o}} )^2 + 4}$, for all $i = 1, \dots , n$, the result follows. 
\hfill$\Box$

\bigskip

In the remainder of this section, we suppress the dependence on $\y $. 
For $i = 1, \dots , n$, let $b_i = - \be _i + \sqrt{{\be _i}^2 + 4}$ and $c_i = ( n_i / 2 + u_i ) \sqrt{{\be _i}^2 + 4}$ and $\ch _i = {c_i}^{1 / 2}$. 
For $i = 1, \dots , n$, let $\uh _i = b_i u_i$, $\vh _i = c_i v_i$, $\vc _i = {\vh _i}^{1 / 2}$, $t_i = \vc _i - 1 / \vc _i$, and $\hat{t} _i = \ch _i t_i$. 
Let $\b = ( b_i )_{i = 1}^{n}$, $\ubh = ( \uh _i )_{i = 1}^{n}$, etc.

\begin{lem}
\label{lem:change} 
\hfill
\begin{itemize}
\item
The conditional density of $( \hat{\t } , \ubh ) | \bbe $ satisfies 
\begin{align}
p( \hat{\t } , \ubh | \bbe ) %
&= \prod_{i = 1}^{n} \Big\{ {- \hat{t} _i / \ch _i + \sqrt{{\hat{t} _i}^2 / {\ch _i}^2 + 4} \over \sqrt{{\hat{t} _i}^2 / {\ch _i}^2 + 4}} {\rm{N}} ( \hat{t} _i | 0, 1) {\rm{Ga}} ( \uh _i | y_i , 1) \Big\} \Big| _{\ch _i = \{ ( n_i / 2 + \uh _i / b_i ) \sqrt{{\be _i}^2 + 4} \} ^{1 / 2}} \non 
\end{align}
for all $( \hat{\t } , \ubh ) \in \mathbb{R} ^n \times (0, \infty )^n$. 
Also, 
\begin{align}
\ubh | \bbe \stackrel{\rm{d}}{=} \ubh \sim \prod_{i = 1}^{n} {\rm{Ga}} ( y_i , 1) \text{.} \non 
\end{align}
\item
For all $g \colon (0, \infty )^n \times (0, \infty )^n \to [0, \infty )$, we have 
\begin{align}
E[ g( \v , \u ) | \bbe ] %
&= E \Big[ g \Big( {1 \over \c } \Big\{ 1 + {\hat{\t } \over \hat{\c }} \Big( {\hat{\t } \over \hat{\c }} + \sqrt{{\hat{\t } ^2 \over \hat{\c } ^2} + 4} \Big) / 2 \Big\} , {\ubh \over \b } \Big) \Big| \bbe \Big] \text{.} \non 
\end{align}
\end{itemize}
\end{lem}

\noindent
{\bf Proof%
.} \ \ We have 
\begin{align}
p( \v , \u | \bbe ) &= \prod_{i = 1}^{n} \Big[ {1 \over \sqrt{2 \pi }} \exp ( c_i ) {1 \over {v_i}^{3 / 2}} \exp \Big\{ - {1 \over 2} \Big( {c_i}^2 v_i + {1 \over v_i} \Big) \Big\} {{b_i}^{y_i} \over \Ga ( y_i )} {u_i}^{y_i - 1} e^{- b_i u_i} \Big] \text{.} \non 
\end{align}
By making the change of variables $\uh _i = b_i u_i$, $i = 1, \dots , n$, 
\begin{align}
&E[ g( \v , \u ) | \bbe ] = \int_{(0, \infty )^n \times (0, \infty )^n} g( \v , \u ) p( \v , \u | \bbe ) d( \v , \u ) \non \\
&= \int_{(0, \infty )^n \times (0, \infty )^n} g \Big( \v , {\ubh \over \b } \Big) \Big( \prod_{i = 1}^{n} \Big[ {1 \over \sqrt{2 \pi }} \exp ( c_i ) {1 \over {v_i}^{3 / 2}} \exp \Big\{ - {1 \over 2} \Big( {c_i}^2 v_i + {1 \over v_i} \Big) \Big\} {1 \over \Ga ( y_i )} {\uh _i}^{y_i - 1} e^{- \uh _i} \Big] \Big) d( \v , \ubh ) \text{.} \non 
\end{align}
By making the change of variables $\vh _i = c_i v_i$, $i = 1, \dots , n$, 
\begin{align}
&E[ g( \v , \u ) | \bbe ] \non \\
&= \int_{(0, \infty )^n \times (0, \infty )^n} g \Big( {\vbh \over \c }, {\ubh \over \b } \Big) \Big( \prod_{i = 1}^{n} \Big[ {1 \over \sqrt{2 \pi }} \exp ( c_i ) {{c_i}^{1 / 2} \over {\vh _i}^{3 / 2}} \exp \Big\{ - {c_i \over 2} \Big( \vh _i + {1 \over \vh _i} \Big) \Big\} {1 \over \Ga ( y_i )} {\uh _i}^{y_i - 1} e^{- \uh _i} \Big] \Big) d( \vbh , \ubh ) \text{.} \non 
\end{align}
By making the change of variables $\vh _i = {\vc _i}^2$, $i = 1, \dots , n$, 
\begin{align}
&E[ g( \v , \u ) | \bbe ] \non \\
&= \int_{(0, \infty )^n \times (0, \infty )^n} g \Big( {\check{\v } ^2 \over \c }, {\ubh \over \b } \Big) \Big( \prod_{i = 1}^{n} \Big[ {1 \over \sqrt{2 \pi }} {2 {c_i}^{1 / 2} \over {\vc _i}^2} \exp \Big\{ - {c_i \over 2} \Big( \vc _i - {1 \over \vc _i} \Big) ^2 \Big\} {1 \over \Ga ( y_i )} {\uh _i}^{y_i - 1} e^{- \uh _i} \Big] \Big) d( \check{\v } , \ubh ) \text{.} \non 
\end{align}
By making the change of variables $t_i = \vc _i - 1 / \vc _i$, or $\vc _i = ( t _i + \sqrt{{t_i}^2 + 4} ) / 2$, $i = 1, \dots , n$, 
\begin{align}
&E[ g( \v , \u ) | \bbe ] \non \\
&= \int_{\mathbb{R} ^n \times (0, \infty )^n} \Big( g \Big( {1 + \t ( \t + \sqrt{\t ^2 + 4} ) / 2 \over \c }, {\ubh \over \b } \Big) \non \\
&\quad \times \Big[ \prod_{i = 1}^{n} \Big\{ {1 \over 2} \Big( 1 + {t_i \over \sqrt{{t_i}^2 + 4}} \Big) {1 \over \sqrt{2 \pi }} {2 {c_i}^{1 / 2} \over 1 + t_i ( t_i + \sqrt{{t_i}^2 + 4} ) / 2} \exp \Big( - {c_i \over 2} {t_i}^2 \Big) {1 \over \Ga ( y_i )} {\uh _i}^{y_i - 1} e^{- \uh _i} \Big\} \Big] \Big) d( \t , \ubh ) \text{.} \non 
\end{align}
By making the change of variables $t_i = \hat{t} _i / \ch _i$, $i = 1, \dots , n$, 
\begin{align}
&E[ g( \v , \u ) | \bbe ] \non \\
&= \int_{\mathbb{R} ^n \times (0, \infty )^n} \Big( g \Big( {1 \over \c } \Big\{ 1 + {\hat{\t } \over \hat{\c }} \Big( {\hat{\t } \over \hat{\c }} + \sqrt{{\hat{\t } ^2 \over \hat{\c } ^2} + 4} \Big) / 2 \Big\} , {\ubh \over \b } \Big) \non \\
&\quad \times \Big[ \prod_{i = 1}^{n} \Big\{ \Big( 1 + {\hat{t} _i / \ch _i \over \sqrt{{\hat{t} _i}^2 / {\ch _i}^2 + 4}} \Big) {1 \over 1 + ( \hat{t} _i / \ch _i ) ( \hat{t} _i / \ch _i + \sqrt{{\hat{t} _i}^2 / {\ch _i}^2 + 4} ) / 2} {\exp (- {\hat{t} _i}^2 / 2) \over \sqrt{2 \pi }} {{\uh _i}^{y_i - 1} e^{- \uh _i} \over \Ga ( y_i )} \Big\} \Big] \Big) d( \hat{\t } , \ubh ) \text{.} \non 
\end{align}
Since 
\begin{align}
\Big( 1 + {t \over \sqrt{t^2 + 4}} \Big) {1 \over 1 + t (t + \sqrt{t^2 + 4} ) / 2} &= \Big( 1 + {t \over \sqrt{t^2 + 4}} \Big) {1 \over 1 + 2 t / (- t + \sqrt{t^2 + 4} )} \non \\
&= \Big( 1 + {t \over \sqrt{t^2 + 4}} \Big) {- t + \sqrt{t^2 + 4} \over t + \sqrt{t^2 + 4}} = {- t + \sqrt{t^2 + 4} \over \sqrt{t^2 + 4}} \non 
\end{align}
for all $t \in \mathbb{R}$, the desired result follows. 
\hfill$\Box$

\bigskip

Let $z \sim {\rm{N}} (0, 1)$. 

\begin{lem}
\label{lem:conditional_expectation} 
For all $i = 1, \dots , n$ and all $h \colon (0, \infty ) \to [0, \infty )$, we have 
\begin{align}
E[ h( \hat{t} _i ) | u_i , \bbe ] &\le 2 E[ h(z) ] \text{.} \non 
\end{align}
\end{lem}

\noindent
{\bf Proof%
.} \ \ Note that 
\begin{align}
{- t + \sqrt{t^2 + 4} \over \sqrt{t^2 + 4}} \le 2 \non 
\end{align}
for all $t \in \mathbb{R}$. 
Then, by Lemma \ref{lem:change}, 
\begin{align}
E[ h( \hat{t} _i ) | u_i , \bbe ] &= \int_{- \infty }^{\infty } h( \hat{t} _i ) {- \hat{t} _i / \ch _i + \sqrt{{\hat{t} _i}^2 / {\ch _i}^2 + 4} \over \sqrt{{\hat{t} _i}^2 / {\ch _i}^2 + 4}} {\rm{N}} ( \hat{t} _i | 0, 1) d{\hat{t} _i} \le \int_{- \infty }^{\infty } h( \hat{t} _i ) 2 {\rm{N}} ( \hat{t} _i | 0, 1) d{\hat{t} _i} \text{,} \non 
\end{align}
which is the desired result. 
\hfill$\Box$

\section{Proof of Theorem \ref{thm:Po}}
\label{sec:proof-geometric} 
In this section, we prove Theorem \ref{thm:Po}. 
We suppress the dependence on $\y $.

\bigskip

\noindent
{\bf Proof of Theorem \ref{thm:Po}.} \ \ Let $M > 0$ be such that $\bPsi + \bPsi ^{- 1} + \X ^{\top } \X \le M \I ^{(p)}$ and such that $M$ is larger than the largest eigenvalue of $\X _{\widetilde{I}} {\X _{\widetilde{I}}}^{\top }$ for all $\emptyset \neq \widetilde{I} \subset \{ 1, \dots , n \} $. 
Fix $0 < \ep < 1$. 

By Lemmas \ref{lem:minorization} and \ref{lem:bound}, the conclusion of the theorem holds if there exist $0 < \de < 1$, $\De > 0$, and $R > 0$ such that for all $\bbe \in \mathbb{R} ^p$ satisfying $\| \bbe \| \ge R$, we have $(P V) ( \bbe ) \le \De + \de V( \bbe )$. 

Let $\bnu = \bPsi \bmu - \X ^{\top } \n $. 
Then, by Lemma \ref{lem:tPop1}, 
\begin{align}
E[ \bbe ^{\top } \bPsi \bbe | \u , \v ] &\le M_1 + \bnu ^{\top } \bOm \bnu + 2 \bnu ^{\top } \bOm \X ^{\top } ( \u + \n / 2) + ( \u+ \n / 2)^{\top } \X \bOm \X ^{\top } ( \u + \n / 2) \text{,} \label{eq:tPop1} 
\end{align}
where $M_1 = M \sum_{k = 1}^{p} ( \bPsi )^{k, k}$ and $\bOm = \{ \bPsi + ( \X ^{\top } \W \X ) \bPsi ^{- 1} ( \X ^{\top } \W \X ) \} ^{- 1}$, for all $\u , \v \in (0, \infty )^n$. 
Fix $\bbe \in \mathbb{R} ^p \setminus \{ \bm{0} ^{(p)} \} $. 
We consider the conditional expectations given $\bbe $ of the terms on the right side of (\ref{eq:tPop1}). 

First, note that $\bOm \le \bPsi ^{- 1}$ for all $\u , \v \in (0, \infty )^n$. 
Then we have 
\begin{align}
E[ \bnu ^{\top } \bOm \bnu | \bbe ] &\le M_2 \text{,} \label{tPop6} 
\end{align}
where $M_2 = \bnu ^{\top } \bPsi ^{- 1} \bnu $. 

Second, by the Schwarz inequality, 
\begin{align}
| \bnu ^{\top } \bOm \X ^{\top } ( \u + \n / 2)| &\le \sqrt{\bnu ^{\top } \bOm \bnu } \sqrt{( \u + \n / 2)^{\top } \X \bOm \X ^{\top } ( \u + \n / 2)} \non \\
&\le \sqrt{\bnu ^{\top } \bPsi ^{- 1} \bnu } \sqrt{( \u + \n / 2)^{\top } \X \bPsi ^{- 1} \X ^{\top } ( \u + \n / 2)} \non 
\end{align}
for all $\u , \v \in (0, \infty )^n$. 
Therefore, 
\begin{align}
E[ | \bnu ^{\top } \bOm \X ^{\top } ( \u + \n / 2)| | \bbe ] &\le \sqrt{\bnu ^{\top } \bPsi ^{- 1} \bnu } \sqrt{E[ ( \u + \n / 2)^{\top } \X \bPsi ^{- 1} \X ^{\top } ( \u + \n / 2) | \bbe ]} \non \\
&\le \sqrt{\bnu ^{\top } \bPsi ^{- 1} \bnu } \sqrt{M^2 E \Big[ \sum_{i = 1}^{n} ( u_i + n_i / 2)^2 \Big| \bbe \Big] } \text{.} \non 
\end{align}
Since 
\begin{align}
E[ ( u_i + n_i / 2)^2 | \bbe ] &= {y_i \over {b_i}^2} + \Big( {y_i \over b_i} + {n_i \over 2} \Big) ^2 \le 3 {{y_i}^2 \over {b_i}^2} + 2 \Big( {n_i \over 2} \Big) ^2 = 3 {y_i}^2 \Big( {\be _i + \sqrt{{\be _i}^2 + 4} \over 4} \Big) ^2 + {{n_i}^2 \over 2} \non \\
&\le 3 {y_i}^2 \Big( {\sqrt{{\be _i}^2 + 4} \over 2} \Big) ^2 + {{n_i}^2 \over 2} = {3 {y_i}^2 \over 4} ( {\be _i}^2 + 4) + {{n_i}^2 \over 2} \non \\
&\le 3 {y_i}^2 + {{n_i}^2 \over 2} + {3 {y_i}^2 \over 4} M \| \bbe \| ^2 \le 3 {y_i}^2 + {{n_i}^2 \over 2} + {3 {y_i}^2 \over 4} M^2 \bbe ^{\top } \bPsi \bbe \text{,} \non 
\end{align}
where $b_i = - \be _i + \sqrt{{\be _i}^2 + 4}$, for all $i = 1, \dots , n$, it follows that 
\begin{align}
E[ | \bnu ^{\top } \bOm \X ^{\top } ( \u + \n / 2)| | \bbe ] &\le \sqrt{\bnu ^{\top } \bPsi ^{- 1} \bnu } \sqrt{M^2 \{ 3 \| \y \| ^2 + \| \n \| ^2 / 2 + (3 \| \y \| ^2 / 4) M^2 \bbe ^{\top } \bPsi \bbe \} } \non \\
&\le \sqrt{\bnu ^{\top } \bPsi ^{- 1} \bnu } M \sqrt{2} \Big\{ \sqrt{3 \| \y \| ^2 + \| \n \| ^2 / 2} + \sqrt{(3 \| \y \| ^2 / 4) M^2} {\bbe ^{\top } \bPsi \bbe \over \sqrt{\bbe ^{\top } \bPsi \bbe }} \Big\} \non \\
&\le M_3 + {M_4 \over \| \bbe \| } \bbe ^{\top } \bPsi \bbe \text{,} \label{tPop7} 
\end{align}
where $M_3 = \sqrt{\bnu ^{\top } \bPsi ^{- 1} \bnu } M \sqrt{2} \sqrt{3 \| \y \| ^2 + \| \n \| ^2 / 2}$ and $M_4 = \sqrt{\bnu ^{\top } \bPsi ^{- 1} \bnu } M \sqrt{2} \sqrt{(3 \| \y \| ^2 / 4) M^2} \sqrt{M}$. 

Third, note that 
\begin{align}
&\x _i ( u_i + n_i / 2) = %
\x _i w_i {\x _i}^{\top } \bbe - \x _i \{ w_i \be _i - ( u_i + n_i / 2) \} \non 
\end{align}
for all $i = 1, \dots , n$ for all $\u , \v \in (0, \infty )^n$. 
Then we have 
\begin{align}
E[ ( \u + \n / 2)^{\top } \X \bOm \X ^{\top } ( \u + \n / 2) | \bbe ] &\le E[ K_1 | \bbe ] + E[ K_2 | \bbe ] + 2 E[ \sqrt{K_1} \sqrt{K_2} | \bbe ] \label{tPop2.5} 
\end{align}
by the Schwarz inequality, where 
\begin{align}
&K_1 = \bbe ^{\top } ( \X ^{\top } \W \X ) \bOm ( \X ^{\top } \W \X ) \bbe \non 
\end{align}
and 
\begin{align}
K_2 &= \Big[ \sum_{i = 1}^{n} \x _i \{ w_i \be _i - ( u_i + n_i / 2) \} \Big] ^{\top } \bOm \Big[ \sum_{i = 1}^{n} \x _i \{ w_i \be _i - ( u_i + n_i / 2) \} \Big] \non \\
&\le \Big[ \sum_{i = 1}^{n} \x _i \{ w_i \be _i - ( u_i + n_i / 2) \} \Big] ^{\top } \bPsi ^{- 1} \Big[ \sum_{i = 1}^{n} \x _i \{ w_i \be _i - ( u_i + n_i / 2) \} \Big] \non \\
&\le n \sum_{i = 1}^{n} [ \x _i \{ w_i \be _i - ( u_i + n_i / 2) \} ]^{\top } \bPsi ^{- 1} [ \x _i \{ w_i \be _i - ( u_i + n_i / 2) \} ] \non \\
&\le n M \sum_{i = 1}^{n} \| \x _i \| ^2 \{ w_i \be _i - ( u_i + n_i / 2) \} ^2 \non 
\end{align}
for $\u , \v \in (0, \infty )^n$. 

Now, let $b_i = - \be _i + \sqrt{{\be _i}^2 + 4}$, $c_i = ( n_i / 2 + u_i ) \sqrt{{\be _i}^2 + 4}$, and $\ch _i = {c_i}^{1 / 2}$ and let $\uh _i = b_i u_i$, $\vh _i = c_i v_i$, $\vc _i = {\vh _i}^{1 / 2}$, $t_i = \vc _i - 1 / \vc _i$, and $\hat{t} _i = \ch _i t_i$ for $i = 1, \dots , n$. 
Let $\b = ( b_i )_{i = 1}^{n}$, $\ubh = ( \uh _i )_{i = 1}^{n}$, etc. 
Let $z \sim {\rm{N}} (0, 1)$. %

By Lemma \ref{lem:change}, 
\begin{align}
E[ K_2 | \bbe ] &\le n M \sum_{i = 1}^{n} \| \x _i \| ^2 K_{2, i} ( \bbe ) \text{,} \non 
\end{align}
where 
\begin{align}
K_{2, i} ( \bbe ) &= E \Big[ \Big| \be _i {\uh _i / b_i + n_i / 2 \over \sqrt{{\be _i}^2 + 4}} \Big\{ 1 + {\hat{t} _i \over \ch _i} \Big( {\hat{t} _i \over \ch _i} + \sqrt{{{\hat{t} _i}^2 \over {\ch _i}^2} + 4} \Big) / 2 \Big\} - ( \uh _i / b_i + n_i / 2) \Big| ^2 \Big| \bbe \Big] \non 
\end{align}
for $i = 1, \dots , n$. 
Fix $i = 1, \dots , n$. 
First, suppose that %
$| \be _i | > \ep \| \bbe \| $. 
Then, by Lemma \ref{lem:conditional_expectation}, 
\begin{align}
&K_{2, i} ( \bbe ) / | \be _i |^2 \non \\
&\le 2 E \Big[ \Big| {\uh _i / b_i + n_i / 2 \over \sqrt{{\be _i}^2 + 4}} - {\uh _i / b_i + n_i / 2 \over \be _i} \Big| ^2 + {( \uh _i / b_i + n_i / 2)^2 \over {\be _i}^2 + 4} {{\hat{t} _i}^2 \over {\ch _i}^2} \Big( {\hat{t} _i \over \ch _i} + \sqrt{{{\hat{t} _i}^2 \over {\ch _i}^2} + 4} \Big) ^2 / 4 \Big| \bbe \Big] \non \\
&\le 2 E \Big[ \Big| {\uh _i / b_i + n_i / 2 \over \sqrt{{\be _i}^2 + 4}} - {\uh _i / b_i + n_i / 2 \over \be _i} \Big| ^2 + {( \uh _i / b_i + n_i / 2)^2 \over {\be _i}^2 + 4} {{\hat{t} _i}^2 \over n_i | \be _i | / 2} \Big\{ {| \hat{t} _i | \over ( n_i | \be _i | / 2)^{1 / 2}} + \sqrt{{{\hat{t} _i}^2 \over n_i | \be _i | / 2} + 4} \Big\} ^2 / 4 \Big| \bbe \Big] \non \\
&\le 2 E \Big[ \Big( {\uh _i \over b_i} + {n_i \over 2} \Big) ^2 \Big| \bbe \Big] \Big[ \Big| {1 \over \sqrt{{\be _i}^2 + 4}} - {1 \over \be _i} \Big| ^2 + {1 \over {\be _i}^2 + 4} 2 E \Big[ {2 z^2 \over n_i | \be _i |} \Big\{ {2^{1 / 2} | z | \over ( n_i | \be _i |)^{1 / 2}} + \sqrt{{2 z^2 \over n_i | \be _i |} + 4} \Big\} ^2 \Big] / 4 \Big] \text{.} \non 
\end{align}
If $\be _i > 0$, then 
\begin{align}
&K_{2, i} ( \bbe ) / | \be _i |^2 \non \\
&\le 2 \Big( 3 {y_i}^2 + {{n_i}^2 \over 2} + {3 {y_i}^2 \over 4} M^2 \bbe ^{\top } \bPsi \bbe \Big) \non \\
&\quad \times \Big[ \Big| {4 \over \be _i \sqrt{{\be _i}^2 + 4} ( \be _i + \sqrt{{\be _i}^2 + 4} )} \Big| ^2 + {1 \over {\be _i}^2 + 4} {1 \over 2} E \Big[ {2 z^2 \over n_i | \be _i |} \Big\{ {2^{1 / 2} | z | \over ( n_i | \be _i |)^{1 / 2}} + \sqrt{{2 z^2 \over n_i | \be _i |} + 4} \Big\} ^2 \Big] \Big] \non \\
&\le 2 \Big( 3 {y_i}^2 + {{n_i}^2 \over 2} + {3 {y_i}^2 \over 4} M^3 \| \bbe \| ^2 \Big) \Big[ {4^2 \over | \be _i |^6} + {1 \over | \be _i |^2} {1 \over 2} E \Big[ {2 z^2 \over n_i | \be _i |} \Big\{ {2^{1 / 2} | z | \over ( n_i | \be _i |)^{1 / 2}} + \sqrt{{2 z^2 \over n_i | \be _i |} + 4} \Big\} ^2 \Big] \Big] \non \\
&\le {2 \over \| \bbe \| } \Big\{ \Big( 3 {y_i}^2 + {{n_i}^2 \over 2} \Big) {1 \over \| \bbe \| ^2} + {3 {y_i}^2 \over 4} M^3 \Big\} \Big[ {4^2 \over \ep ^6 \| \bbe \| ^3} + {1 \over \ep ^2} {1 \over 2} {1 \over n_i \ep } E \Big[ 2 z^2 \Big\{ {2^{1 / 2} | z | \over ( n_i \ep \| \bbe \| )^{1 / 2}} + \sqrt{{2 z^2 \over n_i \ep \| \bbe \| } + 4} \Big\} ^2 \Big] \Big] \text{.} \non 
\end{align}
If $\be _i < 0$, then 
\begin{align}
&K_{2, i} ( \bbe ) / | \be _i |^2 \non \\
&\le 2 E \Big[ \Big( {\uh _i \over 2 | \be _i |} + {n_i \over 2} \Big) ^2 \Big| \bbe \Big] \Big[ \Big( {2 \over | \be _i |} \Big) ^2 + {1 \over | \be _i |^2} {1 \over 2} E \Big[ {2 z^2 \over n_i | \be _i |} \Big\{ {2^{1 / 2} | z | \over ( n_i | \be _i |)^{1 / 2}} + \sqrt{{2 z^2 \over n_i | \be _i |} + 4} \Big\} ^2 \Big] \Big] \non \\
&\le {1 \over 2} E \Big[ \Big( {\uh _i \over \ep \| \bbe \| } + n_i \Big) ^2 \Big| \bbe \Big] \Big[ {4 \over \ep ^2 \| \bbe \| ^2} + {1 \over \ep ^2 \| \bbe \| ^2} {1 \over 2} E \Big[ {2 z^2 \over n_i \ep \| \bbe \| } \Big\{ {2^{1 / 2} | z | \over ( n_i \ep \| \bbe \| )^{1 / 2}} + \sqrt{{2 z^2 \over n_i \ep \| \bbe \| } + 4} \Big\} ^2 \Big] \Big] \non \\
&\le {1 \over 2 \| \bbe \| ^2} E \Big[ \Big( {\uh _i \over \ep \| \bbe \| } + n_i \Big) ^2 \Big] \Big[ {4 \over \ep ^2} + {1 \over \ep ^2} {1 \over 2} E \Big[ {2 z^2 \over n_i \ep \| \bbe \| } \Big\{ {2^{1 / 2} | z | \over ( n_i \ep \| \bbe \| )^{1 / 2}} + \sqrt{{2 z^2 \over n_i \ep \| \bbe \| } + 4} \Big\} ^2 \Big] \Big] \text{,} \non 
\end{align}
where $\uh _i | \bbe \stackrel{\rm{d}}{=} \uh _i \sim {\rm{Ga}} ( y_i , 1)$ by Lemma \ref{lem:change}. 
Next, suppose that $| \be _i | \le \ep \| \bbe \| $. 
Then, by Lemma \ref{lem:conditional_expectation}, 
\begin{align}
K_{2, i} ( \bbe ) %
&\le 2 E \Big[ {\be _i}^2 {( \uh _i / b_i + n_i / 2)^2 \over {\be _i}^2 + 4} \Big\{ 1 + {| \hat{t} _i | \over {n_i}^{1 / 2}} \Big( {| \hat{t} _i | \over {n_i}^{1 / 2}} + \sqrt{{{\hat{t} _i}^2 \over n_i} + 4} \Big) / 2 \Big\} ^2 + ( \uh _i / b_i + n_i / 2)^2 \Big| \bbe \Big] \non \\
&\le 2 E[ ( \uh _i / b_i + n_i / 2)^2 | \bbe ] \Big[ {{\be _i}^2 \over {\be _i}^2 + 4} 2 E \Big[ \Big\{ 1 + {|z| \over {n_i}^{1 / 2}} \Big( {|z| \over {n_i}^{1 / 2}} + \sqrt{{z^2 \over n_i} + 4} \Big) / 2 \Big\} ^2 \Big] + 1 \Big] \non \\
&\le 4 E[ {\uh _i}^2 (1 + \ep ^2 \| \bbe \| ^2 ) + ( n_i / 2)^2 ] \Big[ 2 E \Big[ \Big\{ 1 + {|z| \over {n_i}^{1 / 2}} \Big( {|z| \over {n_i}^{1 / 2}} + \sqrt{{z^2 \over n_i} + 4} \Big) / 2 \Big\} ^2 \Big] + 1 \Big] \text{,} \non 
\end{align}
where the last inequality follows since 
\begin{align}
1 / {b_i}^2 &= \Big( {\be _i + \sqrt{{\be _i}^2 + 4} \over 4} \Big) ^2 \le \Big( {\sqrt{{\be _i}^2 + 4} \over 2} \Big) ^2 = 1 + {{\be _i}^2 \over 4} \le 1 + {\ep ^2 \over 4} \| \bbe \| ^2 \text{.} \non 
\end{align}
Thus, 
\begin{align}
1( \| \bbe \| \ge 1) K_{2, i} ( \bbe ) &\le N \| \bbe \| ^2 \Big( {1 \over \ep ^6 \| \bbe \| } + \ep ^2 \Big) \le M N \Big( {1 \over \ep ^6 \| \bbe \| } + \ep ^2 \Big) \bbe ^{\top } \bPsi \bbe \non 
\end{align}
for all $i = 1, \dots , n$ for some $N > 0$ which depends only on $\n $, $\X $, $\y $, and $M$. 
Hence, 
\begin{align}
E[ K_2 | \bbe ] &\le n M \sum_{i = 1}^{n} \| \x _i \| ^2 M N \Big( {1 \over \ep ^6 \| \bbe \| } + \ep ^2 \Big) \bbe ^{\top } \bPsi \bbe \le n M^2 N \| \X \| ^2 \Big( {1 \over \ep ^6 \| \bbe \| } + \ep ^2 \Big) \bbe ^{\top } \bPsi \bbe \label{tPop3} 
\end{align}
if $\| \bbe \| \ge 1$. 

Note that %
$\X ^{\top } \W \X \ge ( \min_{1 \le i \le n} w_i ) \X ^{\top } \X > \O ^{(p)}$ by assumption. 
Then, by Lemma \ref{lem:tPop2}, 
\begin{align}
K_1 &\le {\tr ( \W ^2 ) \over 1 / M^4 + \tr ( \W ^2 )} \bbe ^{\top } \bPsi \bbe \le \bbe ^{\top } \bPsi \bbe \label{tPop2} 
\end{align}
for all $\u , \v \in (0, \infty )^n$. 
Therefore, 
\begin{align}
E[ \sqrt{K_1} \sqrt{K_2} | \bbe ] &\le \sqrt{\bbe ^{\top } \bPsi \bbe } \sqrt{E[ K_2 | \bbe ]} \le \sqrt{n M^2 N \| \X \| ^2 \Big( {1 \over \ep ^6 \| \bbe \| } + \ep ^2 \Big) } \bbe ^{\top } \bPsi \bbe \label{tPop4} 
\end{align}
if $\| \bbe \| \ge 1$. 
Also, since, by Lemmas \ref{lem:change} and \ref{lem:conditional_expectation}, 
\begin{align}
E [ {w_i}^2 | \bbe ] &= E \Big[ \Big[ {n_i / 2 + \uh _i / b_i \over \sqrt{{\be _i}^2 + 4}} \Big\{ 1 + {\hat{t} _i \over \ch _i} \Big( {\hat{t} _i \over \ch _i} + \sqrt{{{\hat{t} _i}^2 \over {\ch _i}^2} + 4} \Big) / 2 \Big\} \Big] ^2 \Big| \bbe \Big] \non \\
&\le E \Big[ \Big( {n_i / 2 + \uh _i / b_i \over \sqrt{{\be _i}^2 + 4}} \Big) ^2 2 E \Big[ \Big\{ 1 + {|z| \over {n_i}^{1 / 2}} \Big( {|z| \over {n_i}^{1 / 2}} + \sqrt{{z^2 \over n_i} + 4} \Big) / 2 \Big\} ^2 \Big] \Big| \bbe \Big] \non \\
&\le 4 {( n_i / 2)^2 + E[ {\uh _i}^2 | \bbe ] (1 + {\be _i}^2 / 4) \over {\be _i}^2 + 4} E \Big[ \Big\{ 1 + {|z| \over {n_i}^{1 / 2}} \Big( {|z| \over {n_i}^{1 / 2}} + \sqrt{{z^2 \over n_i} + 4} \Big) / 2 \Big\} ^2 \Big] \non \\
&\le 4 {( n_i / 2)^2 + E[ {\uh _i}^2 | \bbe ] \over 4} E \Big[ \Big( 1 + {z^2 \over n_i} + 4 \Big) ^2 \Big] \non 
\end{align}
for all $i = 1, \dots , n$, it follows from Jensen's inequality that 
\begin{align}
E[ K_1 | \bbe ] &\le (1 - \rho ) \bbe ^{\top } \bPsi \bbe \text{,} \label{tPop5} 
\end{align}
where 
\begin{align}
\rho &= 1 / \Big\{ 1 + M^4 \sum_{i = 1}^{n} {( n_i / 2)^2 + E[ {\uh _i}^2 ] \over 1 / E[ (5 + z^2 / n_i )^2 ]} \Big\} \text{.} \non 
\end{align}

By (\ref{tPop2.5}), (\ref{tPop3}), (\ref{tPop4}), and (\ref{tPop5}), 
\begin{align}
&E[ ( \u + \n / 2)^{\top } \X \bOm \X ^{\top } ( \u + \n / 2) | \bbe ] \non \\
&\le \Big\{ (1 - \rho ) + n M^2 N \| \X \| ^2 \Big( {1 \over \ep ^6 \| \bbe \| } + \ep ^2 \Big) + 2 \sqrt{n M^2 N \| \X \| ^2 \Big( {1 \over \ep ^6 \| \bbe \| } + \ep ^2 \Big) } \Big\} \bbe ^{\top } \bPsi \bbe \non \\
&\le \Big\{ (1 - \rho ) + n M^2 N \| \X \| ^2 \Big( {1 \over \ep ^6 \| \bbe \| } + \ep ^2 \Big) + 2 \sqrt{n M^2 N \| \X \| ^2} \sqrt{2} \Big( \sqrt{{1 \over \ep ^6 \| \bbe \| }} + \ep \Big) \Big\} \bbe ^{\top } \bPsi \bbe \non \\
&\le \Big\{ (1 - \rho ) + n M^2 N \| \X \| ^2 \Big( {1 \over \ep ^6 \sqrt{\| \bbe \| }} + \ep \Big) + 2 \sqrt{n M^2 N \| \X \| ^2} \sqrt{2} \Big( {1 \over \ep ^6 \sqrt{\| \bbe \| }} + \ep \Big) \Big\} \bbe ^{\top } \bPsi \bbe \non \\
&\le \Big\{ (1 - \rho ) + M_5 \Big( {1 \over \ep ^6 \sqrt{\| \bbe \| }} + \ep \Big) \Big\} \bbe ^{\top } \bPsi \bbe \text{,} \label{tPop8} 
\end{align}
where $M_5 = n M^2 N \| \X \| ^2 + 2 \sqrt{n M^2 N \| \X \| ^2} \sqrt{2}$, if $\| \bbe \| \ge 1$. 
By (\ref{eq:tPop1}), (\ref{tPop6}), (\ref{tPop7}), and (\ref{tPop8}), 
\begin{align}
(P V) ( \bbe ) &\le M_1 + M_2 + 2 \Big( M_3 + {M_4 \over \| \bbe \| } \bbe ^{\top } \bPsi \bbe \Big) + \Big\{ (1 - \rho ) + M_5 \Big( {1 \over \ep ^6 \sqrt{\| \bbe \| }} + \ep \Big) \Big\} \bbe ^{\top } \bPsi \bbe \non \\
&\le M_1 + M_2 + 2 M_3 + \Big\{ (1 - \rho ) + {2 M_4 + M_5 \over \ep ^6 \sqrt{\| \bbe \| }} + M_5 \ep \Big\} \bbe ^{\top } \bPsi \bbe \non 
\end{align}
if $\| \bbe \| \ge 1$. 
Thus, 
\begin{align}
&(P V) ( \bbe ) \le ( M_1 + M_2 + 2 M_3 ) + (1 - \rho / 3) V( \bbe ) \non 
\end{align}
if 
\begin{align}
&\| \bbe \| \ge 1 + (2 M_4 + M_5 )^2 / ( \ep ^6 \rho / 3) ^2 \text{,} \non 
\end{align}
provided $0 < \ep < 1$ has been chosen so that $\ep \le \rho / (3 M_5 )$. 
This completes the proof. 
\hfill$\Box$

\begin{lem}
\label{lem:tPop1} 
The inequality (\ref{eq:tPop1}) holds. 
\end{lem}

\noindent
{\bf Proof%
.} \ \ Fix $\u , \v \in (0, \infty )^n$. 
First, 
\begin{align}
E[ \bbe ^{\top } \bPsi \bbe | \u , \v ] &= E[ ( \bbe - E[ \bbe | \u , \v ])^{\top } \bPsi ( \bbe - E[ \bbe | \u , \v ]) | \u , \v ] + (E[ \bbe | \u , \v ])^{\top } \bPsi (E[ \bbe | \u , \v ]) \non \\
&\le M \sum_{k = 1}^{p} \Var ( \be _k | \u , \v ) + (E[ \bbe | \u , \v ])^{\top } \bPsi (E[ \bbe | \u , \v ]) \non \\
&\le M \sum_{k = 1}^{p} ( \bPsi )^{k, k} + (E[ \bbe | \u , \v ])^{\top } \bPsi (E[ \bbe | \u , \v ]) \text{.} \non 
\end{align}
Next, 
\begin{align}
(E[ \bbe | \u , \v ])^{\top } \bPsi (E[ \bbe | \u , \v ]) &= \| ( \bPsi ^{1 / 2} )^{\top } ( \bPsi + \X ^{\top } \W \X )^{- 1} \{ \bPsi \bmu + \X ^{\top } ( \u - \n / 2) \} \| ^2 \non \\
&= \| ( \bPsi ^{1 / 2} )^{\top } ( \bPsi + \X ^{\top } \W \X )^{- 1} \{ \bnu + \X ^{\top } ( \u + \n / 2) \} \| ^2 \text{.} \non 
\end{align}
Therefore, 
\begin{align}
&(E[ \bbe | \u , \v ])^{\top } \bPsi (E[ \bbe | \u , \v ]) \non \\
&= \{ \bnu + \X ^{\top } ( \u + \n / 2) \} ^{\top } ( \bPsi + 2 \X ^{\top } \W \X + \X ^{\top } \W \X \bPsi ^{- 1} \X ^{\top } \W \X )^{- 1} \{ \bnu + \X ^{\top } ( \u + \n / 2) \} \non \\
&\le \{ \bnu + \X ^{\top } ( \u + \n / 2) \} ^{\top } \bOm \{ \bnu + \X ^{\top } ( \u + \n / 2) \} \non \\
&= \bnu ^{\top } \bOm \bnu + ( \u + \n / 2)^{\top } \X \bOm \X ^{\top } ( \u + \n / 2) + 2 \bnu ^{\top } \bOm \X ^{\top } ( \u + \n / 2) \non 
\end{align}
and the result follows. 
\hfill$\Box$

\begin{lem}
\label{lem:tPop2} 
The inequalitites in (\ref{tPop2}) hold. 
\end{lem}

\noindent
{\bf Proof%
.} \ \ Fix $\u , \v \in (0, \infty )^n$. 
Then 
\begin{align}
K_1 &= \bbe ^{\top } \{ \bPsi ^{- 1} + ( \X ^{\top } \W \X )^{- 1} \bPsi ( \X ^{\top } \W \X )^{- 1} \} ^{- 1} \bbe \non \\
&= \bbe ^{\top } [ \bPsi - \bPsi \{ \bPsi + ( \X ^{\top } \W \X ) \bPsi ^{- 1} ( \X ^{\top } \W \X ) \} ^{- 1} \bPsi ] \bbe \text{.} \non 
\end{align}
Since 
\begin{align}
( \X ^{\top } \W \X ) \bPsi ^{- 1} ( \X ^{\top } \W \X ) &\le M^2 \X ^{\top } \W ^2 \X \le M^3 \{ \tr ( \W ^2 ) \} \I ^{(p)} \le \{ M^4 \tr ( \W ^2 ) \} \bPsi \text{,} \non 
\end{align}
we have 
\begin{align}
K_1 &\le \bbe ^{\top } ( \bPsi - \bPsi [ \{ 1 + M^4 \tr ( \W ^2 ) \} \bPsi ] ^{- 1} \bPsi ) \bbe = {\tr ( \W ^2 ) \over 1 / M^4 + \tr( \W ^2 )} \bbe ^{\top } \bPsi \bbe \le \bbe ^{\top } \bPsi \bbe \text{.} \non 
\end{align}
This completes the proof. 
\hfill$\Box$

\end{document}